\documentclass[prd, preprint, nofootinbib]{revtex4}
\usepackage[dvipdfmx]{graphicx}
\usepackage{amsfonts}
\usepackage{latexsym}
\usepackage{amssymb}
\usepackage{epsfig}
\usepackage{comment}
\usepackage{here}	
\usepackage{amsmath}

\def\slashb#1{\not\!\!#1}
\newcommand{\im}[1]{\text{Im}\,#1}

\begin{document}

\title{Modular symmetry and non-Abelian discrete flavor symmetries \\
in string compactification}

\author{Tatsuo Kobayashi, Satoshi Nagamoto, Shintaro~Takada, Shio Tamba, and Takuya H. Tatsuishi}
 \affiliation{
Department of Physics, Hokkaido University, Sapporo 060-0810, Japan}



\begin{abstract}
We study the modular symmetry in magnetized D-brane models on $T^2$.
Non-Abelian flavor symmetry $D_4$ in the model with magnetic flux $M=2$ (in a certain unit) is a 
subgroup of the modular symmetry.
We also study the modular symmetry in heterotic orbifold models.
The $T^2/Z_4$ orbifold model has the same modular symmetry as 
the magnetized brane model with $M=2$, and its flavor symmetry $D_4$ is 
a subgroup of the modular symmetry.
\end{abstract}

\preprint{EPHOU-18-003}
\preprint{}

\vspace*{3cm}
\maketitle



\section{Introduction}

Non-Abelian discrete flavor symmetries play an important role in particle physics.
In particular,  many models with various finite groups have been studied in order to explain quark and lepton masses and their mixing angles. 
(See for review \cite{Altarelli:2010gt,Ishimori:2010au,King:2013eh}.)
Those symmetries may be useful for dark matter physics and multi-Higgs models.

Superstring theory is a promising candidate for unified theory including gravity.
It has been shown that some non-Abelian discrete flavor symmetries appear in 
superstring theory with certain compactifications.
Heterotic string theory on toroidal $Z_N$ orbifolds can realize 
non-Abelian flavor symmetries, e.g. $D_4$, and $\Delta (54)$ \cite{Kobayashi:2006wq}.
(See also \cite{Kobayashi:2004ya,Ko:2007dz}.)\footnote{
In Ref.\cite{Beye:2014nxa}, a relation between gauge symmetries and non-Abelian flavor symmetries is discussed 
at the enhancement point.}
Furthermore, magnetized D-brane models within the framework of type II superstring theory 
can lead to similar flavor symmetries \cite{Abe:2009vi,Abe:2009uz,BerasaluceGonzalez:2012vb,Marchesano:2013ega,Abe:2014nla}.
Intersecting D-brane models are T-dual to magnetized D-brane models.
Then, one can realize the same aspects in intersecting D-brane models as in the magnetized ones.\footnote{
See also \cite{Higaki:2005ie}.}

On the other hand, superstring theories on tori and orbifolds have the modular symmetry.
Recently, behavior of zero-modes under modular transformation was studied 
in magnetized D-brane models in Ref. \cite{Kobayashi:2017dyu}.(See also \cite{Cremades:2004wa}.)
Also, behavior of twisted sectors under modular transformation was alreadly studied  
in Ref. \cite{Lauer:1989ax,Lerche:1989cs,Ferrara:1989qb}.
These modular transformations also act non-trivially on flavors and 
transform mutually flavors  each other.
The remarkable difference is that modular transformation also acts Yukawa couplings 
as well as higher order couplings, while those couplings are 
trivial singlets under the usual non-Abelian symmetries.

The purpose of this paper is to study more how modular transformation is represented by 
zero-modes in magnetized D-brane models, 
and to discuss relations between modular transformation and non-Abelian flavor symmetries 
in magnetized D-brane models.
Intersecting D-brane models have the same aspects as magnetized D-brane models, 
because they are T-dual to each other.
Furthermore, 
intersecting D-brane models in type II superstring theory and heterotic string theory have 
similarities, e.g. in two-dimensional conformal field theory.
Thus, here we study modular symmetry and non-Abelian discrete flavor symmetries 
in heterotic orbifold models, too.

This paper is organized as follows.
In section \ref{sec:magne-D}, we study the modular symmetries in magnetized D-brane models and the relation to 
the $D_4$ flavor symmetry.
In section \ref{sec:Het}, we study the modular symmetries in heterotic orbifold models.
Section \ref{sec:conclusion} is conclusion and discussions.
We give brief reviews on non-Abelian discrete flavor symmetries in 
magnetized D-brane models and heterotic orbifold models in 
Appendix  \ref{sec:app-1} and  \ref{sec:app-2}, respectively.

\section{Modular transformation in magnetized D-brane models}
\label{sec:magne-D}

In this section, we study modular transformation of zero-mode wavefunctions in 
magnetized D-brane models.

\subsection{Zero-mode wavefunction}
\label{sec:zero-mode}

Here,  we give a brief review on zero-mode wavefunctions on torus with magnetic flux \cite{Cremades:2004wa}.
For simplicity, we concentrate on $T^2$ with $U(1)$ magnetic flux.
The complex coordinate on $T^2$ is denoted by $z= x^1+ \tau x^2$, where 
$\tau$ is the complex modular parameter, and $x^1$ and $x^2$ are real coordinates.
The metric on $T^2$ is given by 
\begin{equation}
g_{\alpha \beta} = \left(
\begin{array}{cc}
g_{zz} & g_{z \bar{z}} \\
g_{\bar{z} z} & g_{\bar{z} \bar{z}}
\end{array}
\right) = (2\pi R)^2 \left(
\begin{array}{cc}
0 & \frac{1}{2} \\
\frac{1}{2} & 0
\end{array}
\right) .
\end{equation}
We identify $z \sim z +1$ and $z \sim z + \tau$ on $T^2$.

On $T^2$, we introduce the $U(1)$ magnetic flux $F$, 
\begin{equation}
F = i\frac{\pi M}{\im{\tau}}  (dz \wedge d\bar{z}), 
\end{equation}
which corresponds to the vector potential,
\begin{equation}
A(z) = \frac{\pi M}{\im{\tau}} \im{(\bar{z}dz)}.
\end{equation}
Here we concentrate on vanishing Wilson lines.

On the above background, we consider the zero-mode equation for the spinor field 
with the $U(1)$ charge $q=1$,
\begin{equation}
i \slashb{D} \Psi = 0.
\end{equation} 
The spinor field on $T^2$ has two components,  
\begin{equation}
\Psi(z,\bar{z}) = \left(
\begin{array}{c}
\psi_+ \\
\psi_- 
\end{array}
\right).
\end{equation}
The magnetic flux should be quantized such that $M$ is integer.
Either $\psi_+$ or $\psi_-$ has zero-modes exclusively for $M \neq 0$.
For example, we set $M$ to be positive.
Then, $\psi_+$ has $M$ zero-modes, while $\psi_-$ has no zero-mode.
Hence, we can realize a chiral theory.
Their zero-mode profiles are given by 
\begin{equation}
\psi^{j,M}(z) = \mathcal{N} e^{i\pi  M z \frac{\im{z}}{\im{\tau}}} \cdot \vartheta \left[
\begin{array}{c}
\frac{j}{M} \\
0
\end{array}
\right] \left( M z, M\tau \right),
\end{equation}
with $j=0,1,\cdots, (M-1)$, 
where $\vartheta$ denotes the Jacobi theta function, 
\begin{equation}
\vartheta \left[
\begin{array}{c}
a \\
b
\end{array}
\right] (\nu, \tau) = \sum_{l \in {\bf Z}} e^{\pi i (a+l)^2 \tau} e^{2 \pi i (a+l)(\nu+b)} .
\end{equation}
Here, $\mathcal{N}$ denotes  the normalization factor given by 
\begin{equation}
\label{eq:normalization}
\mathcal{N} = \left( \frac{2\im{\tau} M}{\mathcal{A}^2} \right)^{1/4}, 
\end{equation}
with $\mathcal{A}= 4 \pi^2 R^2 \im{\tau}$.

The ground states of scalar fields also have the same profiles as $\psi^{j,M}$.
Thus, the Yukawa coupling including one scalar and two spinor fields can be 
computed by using these zero-mode waverfunctions.
Zero mode wavefunctions satisfy the following relation,
\begin{eqnarray}
\label{eq:psi-psi-psi}
\psi^{i,M}\psi^{j,M} &=& \mathcal{A}^{-1/2} (2\im{\tau})^{1/4}\left(\frac{MN}{M+N}  \right)^{1/4}  \nonumber \\
& & \times \sum_m \psi^{i+j+Mm,M+N} \cdot 
\vartheta \left[
\begin{array}{c}
\frac{Ni-Mj+MNm}{MN(M+N)} \\
0
\end{array}
\right] \left( 0, MN(M+N)\tau \right).
\end{eqnarray}
By use of this relation, 
their Yukawa couplings are given by the wavefunction overlap integral,
\begin{eqnarray}
Y_{ijk} &=& y \int d^2 z \psi^{i,M}\psi^{j,N}(\psi^{k,M'})^{*}  \nonumber \\
 &=& y \left(\frac{2\im{\tau} }{ \mathcal{A}^2}\right)^{1/4} \sum_{m \in Z_{M'}} \delta_{k,i+j+Mm}  
\cdot \vartheta \left[
\begin{array}{c}
\frac{Ni-Mj+MNm}{MNM'} \\
0
\end{array}
\right] \left( 0, MNM'\tau \right),
\end{eqnarray}
where $y$ is constant.
This Yukawa coupling vanishes for $M' \neq M+N$.
Similarly, we can compute higher order couplings 
using the relation (\ref{eq:psi-psi-psi}) \cite{Abe:2009dr}.
In the above equation, the Kronecker delta $\delta_{k,i+j+Mm}$ implies 
the coupling selection rule.
For $g = {\rm gcd}(M,N,M')$,   
non-vanishing Yukawa couplings appear only if 
\begin{equation}\label{eq:selection}
i+j=k \qquad ({\rm mod~} g).
\end{equation}
Hence, we can definite $Z_g$ charges in these couplings \cite{Abe:2009vi}.

\subsection{Modular transformation}

Here, we study modular transformation.
First we give a brief review on results of modular transformation \cite{Kobayashi:2017dyu}.
(See also \cite{Cremades:2004wa}.)
Then, we will study more in details.

The $T^2$ is constructed by ${\mathbb R}^2/\Lambda$, and 
the lattice $\Lambda$ is spanned by the vectors $(\alpha_1,\alpha_2)$, 
where $\alpha_1 = 2\pi R$ and $\alpha_2 = 2 \pi R \tau$.
However, the same lattice can be described by another basis, 
\begin{equation}
\label{eq:SL2Z}
\left(
\begin{array}{c}
\alpha'_2 \\ \alpha'_1
\end{array}
  \right) =\left(
  \begin{array}{cc}
a & b \\
c & d   
\end{array}
\right) \left(
  \begin{array}{c}
\alpha_2 \\ \alpha_1
\end{array}
  \right) ,
\end{equation}
where $a,b,c,d$ are integer with satisfying $ad-bc = 1$.
That is $SL(2,Z)$ transformation.

The modular parameter $\tau = \alpha_2/\alpha_1$ transforms as 
\begin{equation}
\tau \longrightarrow  \frac{a\tau + b}{c \tau + d},
\end{equation}
under (\ref{eq:SL2Z}).
This transformation includes two important generators, $S$ and $T$,
\begin{eqnarray}
& &S:\tau \longrightarrow -\frac{1}{\tau}, \\
& &T:\tau \longrightarrow \tau + 1.
\end{eqnarray}
They satisfy 
\begin{equation}
S^2=\mathbb{I}, \qquad  (ST)^3=\mathbb{I}.
\end{equation}
On top of that, if we impose the algebraic relation, 
\begin{equation}
T^N=\mathbb{I},
\end{equation}
that corresponds to the congruence subgroup of modular group, $\Gamma(N)$.
For example, it is found that $\Gamma(2) \simeq S_3$, $\Gamma(3) \simeq A_4$, $\Gamma(4) \simeq S_4$, 
and  $\Gamma(5) \simeq A_5$.
Since the group $A_4$ is the symmetry of tetrahedron, it is often called the tetrahedral group $T=A_4$.
Also, it may be useful to mention about $\Delta(3N^2) \simeq (Z_N \times Z_N) \rtimes Z_3$ and 
$\Delta(6N^2) \simeq (Z_N \times Z_N) \rtimes S_3$.
We find that $S_3 \simeq \Delta(6)$, $A_4 \simeq \Delta(12)$, and $S_4 \simeq \Delta(24)$.

Following \cite{Kobayashi:2017dyu}, we restrict ourselves to even magnetic fluxes $M$ ($M>0$).
Under $S$, the zero-mode wavefunctions transform as \cite{Cremades:2004wa,Kobayashi:2017dyu}
\begin{equation}
\label{eq:magne-S}
\psi^{j,M} \rightarrow \frac{1}{\sqrt{M}}\sum_k e^{2\pi i jk/M} \psi^{k,M}.
\end{equation}
On the other hand, the zero-mode wavefunctions transform as \cite{Kobayashi:2017dyu}
\begin{equation}
\label{eq:magne-T}
\psi^{j,M} \rightarrow e^{ \pi i j^2/M} \psi^{j,M},
\end{equation}
under $T$.
Generically, the $T$-transformation satisfies 
\begin{equation}
T^{2M} = \mathbb{I},
\end{equation}
on the zero-modes, $\psi^{j,M}$.
Furthermore, in Ref. \cite{Kobayashi:2017dyu} it is shown that 
\begin{equation}
(ST)^3 = e^{\pi i/4},
\end{equation}
on the zero-modes, $\psi^{j,M}$.

In what follows, we study more concretely.

\subsubsection{Magnetic flux $M=2$}

Let us study the case with the magnetic flux $M=2$ concretely.
There are two zero-modes, $\psi^{0,2}, \psi^{1,2}$.
The $S$-transformation acts on these zero-modes as 
\begin{equation}
\left( \begin{array}{c}
 \psi^{0,2} \\ 
\psi^{1,2} 
\end{array} \right) \longrightarrow S_{(2)}\left( \begin{array}{c}
 \psi^{0,2} \\ 
\psi^{1,2} 
\end{array} \right), \qquad S_{(2)} = \frac{1}{\sqrt 2}\left(
\begin{array}{cc}
1  & 1 \\
1  &  -1
\end{array}
\right).
\end{equation} 
The $T$-transformation acts as 
\begin{equation}
\left( \begin{array}{c}
 \psi^{0,2} \\ 
\psi^{1,2} 
\end{array} \right) \longrightarrow T_{(2)}\left( \begin{array}{c}
 \psi^{0,2} \\ 
\psi^{1,2} 
\end{array} \right), \qquad T_{(2)} = \left(
\begin{array}{cc}
1  & 0 \\
0  &  i
\end{array}
\right).
\end{equation} 
They satisfy the following algebraic relations,
\begin{equation}
\label{eq:ST-M=2}
S^2_{(2)}=\mathbb{I}, \qquad T^4_{(2)}=\mathbb{I}, \qquad (S_{(2)}T_{(2)})^3=e^{\pi i /4}\mathbb{I}.
\end{equation}
They construct a closed algebra with the order 192, which we denote here 
by $G_{(2)}$.
By such an algebra, modular transformation is represented by two zero-modes, $\psi^{0,2}, \psi^{1,2}$.
We find that $(S_{(2)}T_{(2)})^3$ is a center.
Indeed, there are eight center elements and their group is $Z_8$.
Other diagonal elements correspond to $Z_4$, which is generated by $T_{(2)}$.
Here, we denote 
\begin{equation}
a = (S_{(2)}T_{(2)})^3, \qquad   a' = T_{(2)}.
\end{equation}
The diagonal elements are represented by $a^ma'^n$, i.e. $Z_8 \times Z_4$.

Here, we examine the right coset $Hg$ for $g \in G_{(2)}$, where $H$ is the above $Z_8 \times Z_4$, i.e. 
$H=\{ a^ma'^n  \}$.
There would be $6(=192/(8\times 4))$ cosets.
Indeed, we obtain the following six cosets:
\begin{equation}
H, \quad HS_{(2)}, \quad HS_{(2)}T^k_{(2)}, \quad HS_{(2)}T^2_{(2)}S_{(2)},
\end{equation}
with $k=1,2,3$.
By simple computations, we find $HS_{(2)}T^k_{(2)}S_{(2)} \sim HS_{(2)}T^{4-k}_{(2)}S_{(2)} $ 
and $HS_{(2)}T^2_{(2)}S_{(2)}T \sim HS_{(2)}T^2_{(2)}S_{(2)}$.

Furthermore, we would make a (non-Abelian) subgroup with the order 6 
by choosing properly six elements such that we  
pick one element up from 
each coset and their algebra is closed.
The non-Abelian group with the order 6 is unique, i.e. $S_3$.
For example, we may be able to obtain the $Z_3$ generator from $ HS_{(2)}T_{(2)}$  
because $(S_{(2)} T_{(2)})^3 \in H$.
That is, we define 
\begin{equation}
b=a^ma'^nS_{(2)}T_{(2)}.
\end{equation}
Then, we require $b^3 = \mathbb{I}$.
There are three solutions, $(m,n)=(3,2)$, (5,0)  mod (8,4).
Similarly, we can obtain the $Z_2$ generator e.g. form $HS_{(2)}T_{(2)}^2S_{(2)}$ 
because   $(S_{(2)}T_{(2)}^2S_{(2)})^2 \in H$.
Then, we define
\begin{equation}
c=a^{m'}a'^{n'}S_{(2)}T_{(2)}^2S_{(2)}.
\end{equation}
We find $c^2 = \mathbb{I}$ when $n'=-m'$ mod 4.
On top of that, we require $(bc)^2 = \mathbb{I}$, and that leads to the 
conditions, $n=-m'-1$ mod 4 and $m=m' + 2$ mod 8.
As a result, there are six solutions, 
$(m,n,m')=(3,2,1)$, (3,2,5), (5,0,3), (5,0,7) with $n'=-m'$ mod 4.

For example, for $(m,n,m')=(3,2,5)$ we write 
\begin{equation}
b = \frac{1}{\sqrt 2}\left(
\begin{array}{cc}
 \rho^3 & \rho^{-3} \\
 \rho^{-1} & \rho^{-3}
 \end{array}\right), \qquad 
 c = \left(
\begin{array}{cc}
 0 & \rho^{-3} \\
 \rho^3 & 0
 \end{array}\right).
\end{equation}
The six elements of the subgroup are written explicitly, 
\begin{eqnarray}
& & \left(
\begin{array}{cc}
1 & 0 \\
0 & 1 
\end{array} \right), \qquad 
\frac{1}{\sqrt 2}\left(
\begin{array}{cc}
 1 & 1 \\
 1 & -1
 \end{array}\right), \qquad 
\left(
\begin{array}{cc}
 0 & \rho^{-3} \\
 \rho^3 & 0
 \end{array}\right),    \nonumber \\
& & \frac{1}{\sqrt 2}\left(
\begin{array}{cc}
-1 & i \\
-i & 1 
\end{array} \right), \qquad 
\frac{1}{\sqrt 2}\left(
\begin{array}{cc}
 \rho^3 & \rho^{-3} \\
 \rho^{-1} & \rho^{-3}
 \end{array}\right), \qquad 
\frac{1}{\sqrt 2}\left(
\begin{array}{cc}
 \rho^{-3} & \rho \\
 \rho^3 & \rho^3
 \end{array}\right),   
\end{eqnarray}
where $\rho = e^{2\pi i /8}$.
They correspond to $S_3\simeq \Gamma(2) \simeq \Delta(6)$ because 
they satisfy the following algebraic relations, 
\begin{equation}
c^2 = b^3 = (bc)^2=\mathbb{I}.
\end{equation}
Moreover, they satisfy the following algebraic relation with $Z_8\times Z_4$,
\begin{equation}
b^{-1}ab^{1}= a,\qquad cac=a,\qquad 
b^{-1}a'b=a, \qquad
ca'c^{-1} = a^2a'^3. 
\end{equation}
Thus, the algebra of $G_{(2)}$ is isomorphic to 
$(Z_8 \times Z_4) \rtimes S_3$.

We have started by choosing $HS_{(2)}T_{(2)}^2S_{(2)}$ for a candidate of the
$Z_2$ generator.
We can obtain the same results by starting with $HS_{(2)}$ for 
a candidate of the $Z_2$ generator.

\subsubsection{Magnetic flux $M=4$}

Similarly, we study the case with the magnetic flux $M=4$.
There are four zero-modes, $\psi^{i,M}$ with $i=0,1,2,3$.
The $S$ and $T$-transformations are  represented by  $\psi^{i,M}$,
\begin{equation}
S_{(4)} = \frac12\left(
\begin{array}{cccc}
1 & 1 &  1 & 1 \\
1 & i &  -1& -i \\
1 & -1 & 1 & -1 \\
1 &-i & -1 & i
\end{array}
\right),   \qquad 
T_{(4)} = \left(
\begin{array}{cccc}
1 &  &   &  \\
 & e^{\pi i/4} &  &  \\
 &  & -1 &  \\
 & &  & e^{\pi i/4}
\end{array}
\right).
\end{equation}
This is a reducible representation.
In order to obtain irreducible representations, we use the flowing basis,
\begin{equation}
\left(  
\begin{array}{c}
\psi^{0.4} \\  \psi^{1,4}_{+} \\ \psi^{2,4} 
\end{array}
 \right) = \left(  
\begin{array}{c}
\psi^{0.4} \\  \frac{1}{\sqrt{2}}(\psi^{1,4} + \psi^{3,4}) \\ \psi^{2,4} 
\end{array}
 \right), \qquad 
 \psi^{1,4}_- = \frac{1}{\sqrt{2}}(\psi^{1,4} - \psi^{3,4}).
\end{equation}
This is nothing but zero-modes on the $T^2/Z_2$ orbifold \cite{Abe:2008fi}.
The former corresponds to $Z_2$ even states, while the latter 
corresponds to the $Z_2$ odd state.
Note that $(ST)^3$ transforms the lattice basis $(\alpha_1,\alpha_2) \rightarrow (-\alpha_1,-\alpha_2)$.
Thus, it is reasonable that the zero-modes on the  $T^2/Z_2$ orbifold correspond to 
the irreducible representations.

The $S$ and $T$-representations by the $Z_2$ odd zero-mode are quite simple, 
and these are represented by 
\begin{equation}
S_{(4-)} = i, \qquad T_{(4)-} = e^{\pi i/4}.
\end{equation}
Their closed algebra is $Z_8$.

On the other hand,  the $S$ and $T$-transformations are represented by the $Z_2$ even zero-modes,
\begin{equation}
S_{(4)+}=\frac12\left(
\begin{array}{ccc}
1 & \sqrt2 & 1 \\
\sqrt 2 & 0 & -\sqrt 2 \\
1 & -\sqrt 2 & 1
\end{array}
\right), \qquad T_{(4)+}=\left(
\begin{array}{ccc}
1 & & \\
  & e^{\pi i/4} & \\
  & & -1
  \end{array}\right).
  \end{equation}
They satisfy the following algebraic relation,
\begin{equation}
(S_{(4)+})^2= \mathbb{I},\qquad (T_{(4)+})^8=\mathbb{I}, \qquad (S_{(4)+}T_{(4)+})^3=e^{\pi i/4} \mathbb{I}.
\end{equation}
We denote the closed algebra of $S_{(4)+}$ and $T_{(4)+}$ by $G_{(4)+}$.
Its order is equal to 768, and it includes the center element $(S_{(4)+}T_{(4)+})^3$, 
i.e. $Z_8$.
Other diagonal elements correspond to $Z_8$, which is generated by $T_{(4)+}$.
Again, we denote $a=(S_{(4)+}T_{(4)+})^3$ and $a'=T_{(4)+}$, 
and the diagonal elements are written by $a^ma'^n$, i.e. $Z_8 \times Z_8$.

Similar to the case with $M=2$, we examine the coset structure, $Hg$.
Indeed, there are the following 12 cosets:
\begin{equation}
H, \quad HS_{(4)+}, \quad HS^k_{(4)+}, \quad HS_{(4)+}T^\ell_{(4)+} S_{(4)+},
\end{equation}
where $k=1,\cdots,7$ and $\ell = 2,4,6$.
By simple computation, we find that 
\begin{equation}
HS_{(4)+}T^k_{(4)+} S_{(4)+} \sim HS_{(4)+}T^{8-k}_{(4)+}, \qquad 
HS_{(4)+}T^\ell_{(4)+} S_{(4)+}T \sim HS_{(4)+}T^{8-\ell}_{(4)+} S_{(4)+}
\end{equation}
for $k=$ odd and $\ell=$ even.

We make a subgroup with the order 12 by choosing properly 12 elements 
such that we pick one element up from each coset and their algebra is 
closed.
The non-Abelian group with the order 12 are $D_6$, $Q_6$ and $A_4$.
Among them, $A_4$ would be a good candidate. 
Indeed, we can obtain the $Z_3$ generator from $HS_{(4)+}T_{(4)+}$, gain.
That is, we define
\begin{equation}
t=a^ma'^nS_{(4)+}T_{(4)+}.
\end{equation}
The solutions for $t^3=\mathbb{I}$ are obtained by $(m,n)=(1,4)$, (3,6), (5,0),  and (7,2).
We also define 
\begin{equation}
s=a^{m'}a'^{n'}S_{(4)+}T^4_{(4)+}S_{(4)+}.
\end{equation}
The solutions for $s^2=\mathbb{I}$ are obtained by $(m',n')=(0,0)$, (0,4), (4,0) and (4,4).
These two generators satisfy $(st)^3=\mathbb{I}$ if $(m',n')=(0,4)$, and (4,0), i.e. 
\begin{equation}
s=\left(
\begin{array}{ccc}
0 & 0& \pm 1 \\
0 & -1 & 0 \\
\pm 1 & 0 &0
\end{array}\right).
\end{equation}
As a result, they satisfy 
\begin{equation}
s^2=t^3=(st)^3=\mathbb{I}.
\end{equation}
That is the $A_4$ algebra.

\subsubsection{Large magnetic flux $M$}

For larger magnetic fluxes, $S$ and $T$-transformations are represented by 
zero-modes $\psi^{j,M}$, but those are reducible representations.
The irreducible representations are obtained in the $T^2/Z_2$ orbifold basis, 
\begin{equation}
\psi_{\pm}^{j,M} = \frac{1}{\sqrt 2}\left(  \psi^{j,M} \pm \psi^{M-j,j} \right).
\end{equation}
The representations of $T_{(M)}$ are simply obtained by
\begin{equation}
T_{(M)+}\left(
\begin{array}{c}
\psi^{0,M} \\
\psi^{1,M}_+ \\
\vdots \\
\psi^{j,M}_+ \\
\vdots\\
\psi^{M/2,M}
\end{array} \right) = 
\left(
\begin{array}{cccccc}
1  & & & & & \\
 & e^{\pi i/M} & & & & \\
 & & \ddots & & & \\
 & & & e^{\pi i j^2/M} & & \\
 & & & & \ddots & \\
 & & & & &  e^{\pi i M/4}
\end{array}\right)
\left(
\begin{array}{c}
\psi^{0,M} \\
\psi^{1,M}_+ \\
\vdots \\
\psi^{j,M}_+ \\
\vdots\\
\psi^{M/2,M} 
\end{array} \right),
\end{equation}
and 
\begin{equation}
T_{(M)-}\left(
\begin{array}{c}
\psi^{1,M}_- \\
\vdots \\
\psi^{j,M}_- \\
\vdots\\
\psi^{M/2-1,M}_- 
\end{array} \right) = 
\left(
\begin{array}{ccccc}
  e^{\pi i/M} & & & & \\
  & \ddots & & & \\
  & & e^{\pi i j^2/M} & & \\
  & & & \ddots & \\
  & & & &  e^{\pi (M/2-1)^2/M}
\end{array}\right)
\left(
\begin{array}{c}
\psi^{1,M}_- \\
\vdots \\
\psi^{j,M} \\
\vdots\\
\psi^{M/2-1,M}_- 
\end{array} \right).
\end{equation}
Both correspond to $Z_{2M}$.

On the other hand, the $S_{(M)\pm}$ transforms 
\begin{equation}
S_{(M)\pm}\psi^{j,M}_{\pm}  = \frac{1}{\sqrt {2M}}\sum_k 
\left(e^{2 \pi jk/M}\pm e^{2 \pi i (M-j)k/M,M} \right)\psi^{k,M} .
\end{equation}
This representation is also written by
\begin{equation}
S_{(M)\pm}\psi^{j,M}_{\pm}  = \frac{1}{\sqrt {2M}}\sum_k 
\left(e^{2 \pi (M-j)(M-k)/M}\pm e^{2 \pi i j(M-k)/M,M} \right)\psi^{M-k,M} .
\end{equation}
Thus, the $S$-transformation is represented on the $T^2/Z_2$ orbifold basis by
\begin{equation}
S_{(M)\pm}\psi^{j,M}_{\pm}  = \frac{1}{\sqrt {M}}\sum_{k \leq M/2 }
\left(e^{2 \pi jk/M}\pm e^{2 \pi i (M-j)k/M} \right)\psi^{k,M}_{\pm} .
\end{equation}
These are written by 
\begin{eqnarray}
S_{(M)+}\psi^{j,M}_{+}  = \frac{2}{\sqrt {M}}\sum_{k \leq M/2 } \cos (2 \pi jk/M) \psi^{j,M}_{+}, \nonumber \\
S_{(M)-}\psi^{j,M}_{-}  = \frac{2i}{\sqrt {M}}\sum_{k \leq M/2 } \sin (2 \pi jk/M) \psi^{j,M}_{-}. 
\end{eqnarray}

For example, for $M=6$, $S$ and $T$ are represented by $Z_2$ even zero-modes, 
\begin{equation}
S_{(6)+}\left(
\begin{array}{c}
\psi^{0,6}   \\
\psi^{1,6}_+ \\
\psi^{2,6}_+ \\
\psi^{3,6}
\end{array}\right) = \frac{1}{\sqrt 6}\left(
\begin{array}{cccc}
1 &\sqrt 2 & \sqrt 2 & 1 \\
\sqrt 2 & 1  & -1 & -\sqrt 2 \\
\sqrt 2 & -1 & -1 &  \sqrt 2 \\
 1& -\sqrt 2 & \sqrt 2 & -1
\end{array} \right) \left(
\begin{array}{c}
\psi^{0,6}   \\
\psi^{1,6}_+ \\
\psi^{2,6}_+ \\
\psi^{3,6}
\end{array}\right),
\end{equation}
\begin{equation}
T_{(6)+}\left(
\begin{array}{c}
\psi^{0,6}   \\
\psi^{1,6}_+ \\
\psi^{2,6}_+ \\
\psi^{3,6}
\end{array}\right) = \left(
\begin{array}{cccc}
1 & & & \\
& e^{\pi i /6}  & & \\
 & &  e^{2 \pi i/3} &   \\
 & & & e^{^{3 \pi i /2}}
\end{array} \right) \left(
\begin{array}{c}
\psi^{0,6}   \\
\psi^{1,6}_+ \\
\psi^{2,6}_+ \\
\psi^{3,6}
\end{array}\right),
\end{equation}
while $S$ and $T$ are represented by $Z_2$ odd zero-mode,
\begin{equation}
S_{(6)-}\left(
\begin{array}{c}
\psi^{1,6}_- \\
\psi^{2,6}_-
\end{array}\right) = \frac{i}{ \sqrt 2}\left(
\begin{array}{cc}
1 & 1 \\
 1 & -1
\end{array} \right) \left(
\begin{array}{c}
\psi^{1,6}_- \\
\psi^{2,6}_-
\end{array}\right),
\end{equation}
\begin{equation}
T_{(6)-}\left(
\begin{array}{c}
\psi^{1,6}_- \\
\psi^{2,6}_-
\end{array}\right) = \left(
\begin{array}{cc}
e^{\pi i /6} & 0 \\
 0 & e^{2 \pi i/3}
\end{array} \right) \left(
\begin{array}{c}
\psi^{1,6}_- \\
\psi^{2,6}_-
\end{array}\right).
\end{equation}

\subsection{Non-Abelian discrete flavor symmetries}
\label{sec:D4}

In Ref.\cite{Abe:2009vi},  it is shown that the models with $M=2$ as well as 
even magnetic fluxes have the $D_4$ flavor symmetry.
See Appendix \ref{sec:app-1}.
One of the $Z_2$ elements in $D_4$ corresponds to $(T_{(2)})^2$ on the zero-modes, $\psi^{0,2}$ and $\psi^{1,2}$,
i.e. 
\begin{equation}
Z = \left(
\begin{array}{cc}
1 & 0 \\
0 & -1
\end{array}
\right) = (T_{(2)})^2.
\end{equation}
In addition, the permutation $Z^C_2$ element in $D_4$ corresponds to 
$S_{(2)}T_{(2)}T_{(2)}S_{(2)}$, i.e.
\begin{equation}
C = \left(
\begin{array}{cc}
0 & 1 \\
1 & 0
\end{array}
\right) = S_{(2)}T_{(2)}T_{(2)}S_{(2)}.
\end{equation}
Thus, the $D_4$ group, which includes the eight elements (\ref{eq:D4-8}), is 
subgroup of $G_{(2)} \simeq (Z_8 \times Z_4) \rtimes S_3$.

However, there is the difference between the modular symmetry and 
the $D_4$ flavor symmetry, which studied in  Ref.\cite{Abe:2009vi}.
The modular symmetry transforms the Yukawa couplings, while 
the Yukawa couplings are invariant under the $D_4$ flavor symmetry.
In order to study this point, here we examine the Yukawa couplings 
among $\psi^{i,2}$, $\psi'^{j,2}$ and $\psi^{k,4}$.
Both $\psi^{i,2}$ and $\psi'^{j,2}$ are $D_4$ doublets, and 
their tensor product ${\bf 2}\times {\bf 2}$ is expanded by
\begin{equation}
{\bf 2} \times {\bf 2} = {\bf 1}_{++} + {\bf 1}_{+-} + {\bf 1}_{-+} + {\bf 1}_{--}.
\end{equation}
Thus, the products  $\psi^{i,2} \psi'^{j,2}$ correspond to four singlets,
\begin{eqnarray}
& & {\bf 1}_{+\pm}: \psi^{0,2} \psi'^{0,2} \pm \psi^{1,2} \psi'^{1,2}, \qquad 
 {\bf 1}_{-\pm}: \psi^{0,2} \psi'^{1,2} \pm \psi^{1,2} \psi'^{0,2}.
\end{eqnarray}

On the other hand, by use of Eq.(\ref{eq:psi-psi-psi}), 
the products $\psi^{i,2}\psi'^{j,2}$ are expanded by $\psi^{k,4}$.
For example, we can expand as 
\begin{eqnarray}
\psi^{0,2} \psi'^{0,2} \pm \psi^{1,2} \psi'^{1,2} &\sim & \left( 
Y^{(0)}( 16\tau ) + Y^{(8/16)}( 16\tau ) \pm \left(  Y^{(4/16)}( 16\tau ) + Y^{(12/16)}( 16\tau )    \right)
\right) \nonumber \\  & & \times \left( \psi^{0,4} \pm \psi^{2,4} \right)
\end{eqnarray}
up to constant factors, where 
\begin{equation}
Y^{(j/M)}(M\tau) = \mathcal{N}  \cdot \vartheta \left[
\begin{array}{c}
\frac{j}{M} \\
0
\end{array}
\right] \left( 0, M\tau \right).
\end{equation}
Note that $Y^{(j/M)}(M\tau) = Y^{(1-j/M)}(M\tau)$.
It is found that 
\begin{eqnarray}
 & & (T_{(4)})^2 \left( \psi^{0,4} \pm \psi^{2,4} \right) = \left( \psi^{0,4} \pm \psi^{2,4} \right),   \nonumber \\
 & & (S_{(4)}T_{(4)}T_{(4)}S_{(4)}) \left( \psi^{0,4} \pm \psi^{2,4} \right) = \pm \left( \psi^{0,4} \pm \psi^{2,4} \right).
\end{eqnarray}
Thus, the zero-modes $\psi^{0,4} \pm \psi^{2,4} $ are indeed $D_4$ singlets, ${\bf 1}_{+\pm}$ 
when we identify $(T_{(4)})^2 $ and $(S_{(4)}T_{(4)}T_{(4)}S_{(4)})$ as $Z_2$ and $Z^C_2$ of $D_4$.
In this sense, the $D_4$ flavor symmetry is a subgroup of the modular symmetry.
Also, it is found that the above Yukawa couplings, $Y^{(m/4)}(16\tau)$, with $m=0,1,2,3$ are 
invariant under $T^2$ and $STTS$ transformation.

Similarly, we can expand 
\begin{eqnarray}\label{eq:01-13}
\psi^{0,2} \psi'^{1,2} + \psi^{1,2} \psi'^{0,2} &\sim & \left( 
Y^{(2/16)}( 16\tau ) + Y^{(6/16)}( 16\tau )  
\right)   \times \left( \psi^{1,4} + \psi^{3,4} \right),
\end{eqnarray}
up to constant factors.
It is found that 
\begin{eqnarray}
 & & (T_{(2)})^2 \left( \psi^{0,2} \psi'^{1,2} + \psi^{1,2} \psi'^{0,2} \right) =   - \left( \psi^{0,2} \psi'^{1,2} + \psi^{1,2} \psi'^{0,2} \right) .
\end{eqnarray}
On the other hand, we obtain 
\begin{eqnarray}
 & & (T_{(4)})^2 \left( \psi^{1,4} + \psi^{3,4} \right) =   i  \left( \psi^{1,4} + \psi^{3,4} \right) .
\end{eqnarray}
In addition, we find 
\begin{equation}
T^2: \left( 
Y^{(2/16)}( 16\tau ) + Y^{(6/16)}( 16\tau )  
\right)   \rightarrow i \left( 
Y^{(2/16)}( 16\tau ) + Y^{(6/16)}( 16\tau )  
\right)  .
\end{equation}
Thus, the $T^2$ transformation is consistent between left and right hand sides in (\ref{eq:01-13}).
However, when we interpret $T^2$ as $Z_2$ of the $D_4$ flavor symmetry, 
we face inconsistency, because Yukawa couplings are not invariant and 
$ (\psi^{1,4} + \psi^{3,4} )$ has transformation behavior different from 
$(\psi^{0,2} \psi'^{1,2} + \psi^{1,2} \psi'^{0,2})$.
We can make this consistent by defining 
$Z_2$ of the $D_4$ on $ (\psi^{1,4} + \psi^{3,4} )$  such that 
its transformation absorbs the phase of Yukawa couplings under $T^2$ transformation.
Then, the mode  $ (\psi^{1,4} + \psi^{3,4} )$ exactly corresponds to the 
$D_4$ singlet, ${\bf 1}_{-+}$.
We find that $(\psi^{0,2} \psi'^{1,2} + \psi^{1,2} \psi'^{0,2})$ is invariant under $S_{(2)}T_{(2)}T_{(2)}S_{(2)}$, 
and $ (\psi^{1,4} + \psi^{3,4} )$ is also invariant under $S_{(4)}T_{(4)}T_{(4)}S_{(4)}$.
That is consistent.
Therefore, the $D_4$ flavor symmetry is a subgroup of the modular symmetry on $\psi^{j,2}$ $(j=0,1)$.
However, when the model includes couplings to zero-modes with larger $M$, 
we have to modify their modular symmetries such that coupling constants are invariant under the flavor symmetry.
Then, we can define the $D_4$ flavor symmetry.

Here, we give a comment on the $T^2/Z_2$ orbifold.
The $T^2/Z_2$ orbifold basis gives the irreducible representations of the modular symmetry.
The $D_4$ flavor symmetry is defined through the modular symmetry, as above.
That is the reason why the $D_4$ flavor symmetry remains on the $T^2/Z_2$ orbifold  \cite{Marchesano:2013ega,Abe:2014nla}.

\section{Heterotic orbifold models}
\label{sec:Het}

Intersecting D-brane models in type II superstring theory is 
T-dual to magnetized D-brane models.
Thus, intersecting D-brane models also have the same behavior under modular transformation 
as magnetized D-brane models.
Furthermore, intersecting D-brane models in type II superstring theory 
and heterotic string theory on orbifolds have similarities, e.g. 
in two-dimensional conformal field theory.
For example, computations of 3-point couplings as well as $n$-point couplings are similar to 
each other.
Here, we study modular symmetry in heterotic orbifold models.
Using results in Ref.~\cite{Lauer:1989ax,Lerche:1989cs,Ferrara:1989qb}, 
we compare the modular symmetries in heterotic orbifold models with non-Abelian flavor symmetries and also
the modular symmetries in the magnetized D-brane models, which have been derived in the previous section.

\subsection{Twisted sector}

Here, we give a brief review on heterotic string theory on orbifolds.
The orbifold is the division of the torus $T^n$ by the $Z_N$ twist $\theta$, i.e. 
$T^n/Z_N$.
Since the $T^n$ is constructed by ${\mathbb R}^n/\Lambda$, the $Z_N$ twist $\theta$ 
should be an automorphism of the lattice $\Lambda$.
Here, we focus on two-dimensional orbifolds, $T^2/Z_N$.
The six-dimensional orbifolds can be constructed by products of two-dimensional ones.
All of the possible orbifolds are classified as $T^2/Z_N$ with $N=2,3,4,6$.

On orbifolds, there are fixed points, which satisfy the 
following condition,
\begin{equation}
\label{eq:fp}
x^i = (\theta^n x)^i + \sum_k m_k \alpha_k^i,
\end{equation}
where $x^i$ are real coordinates, $\alpha_k^i$ are two lattice vectors, 
and $m_k$ are integer for $i,k=1,2$.
Thus, the fixed points can be represented by corresponding space group 
elements $(\theta^n, \sum_k m_k \alpha_k^i)$, or in short $(\theta^n, ( m_1,m_2 ))$.

The heterotic string theory on orbifolds has localized modes at fixed points, 
and these are the so-called twisted strings.
These twisted states can be labeled by use of fixed points, $\sigma_{\theta,(m_1,m_2)}$.
All of the twisted states $\sigma_{\theta,(m_1,m_2)}$   have the same spectrum, if 
discrete Wilson lines vanish.
Thus, the massless modes are degenerate by the number of fixed points.

On the $T^2/Z_2$ orbifold, there are four fixed points, which are denoted by 
$(\theta,(0,0))$,  $(\theta,(1,0))$,  $(\theta,(0,1))$,  $(\theta,(1,1))$.
The corresponding twisted states are denoted by 
$\sigma_{\theta,(m,n)}$ for $m,n=0,1$.

On the   $T^2/Z_3$ orbifold, $\alpha_1$ and $\alpha_2$ correspond to 
the $SU(3)$ simple roots and they are identified each other by the $Z_3$ twist.
Thus, three fixed points on the $T^2/Z_3$ orbifold are represented by 
the space group elements, $(\theta,m \alpha_1)$ for $m=0,1,2$, 
or in short $(\theta, m)$.
The corresponding twisted states are denoted by 
$\sigma_{\theta,m}$ for $m=0,1,2$.

Similarly, we can obtain the fixed points and twisted states 
on the $T^2/Z_4$, where 
$\alpha_1$ and $\alpha_2$ correspond to the $SO(4)$ simple roots and 
they are identified each other by the $Z_4$ twist.
For the $Z_4$ twist $\theta$,  two fixed points satisfy Eq.(\ref{eq:fp}), 
and these can be represented by $(\theta, m\alpha_1)$ for $m=0,1$, or in short 
$(\theta, m)$.
Then, the first twisted states are denoted by 
$\sigma_{\theta,m}$ for $m=0,1$.
In addition, for $\theta^2$, there are four points, which 
satisfy  Eq.(\ref{eq:fp}), and these can denoted by 
$(\theta^2,(m,n))$ for  $m,n=0,1$.
Indeed, these correspond to the four fixed points on the $T^2/Z_2$ orbifold.
Then, the second twisted states are denoted by 
$\sigma_{\theta^2,(m,n)}$ for $m,n=0,1$.
However, the fixed points $(\theta^2,(1,0))$ and $(\theta^2,(0,1))$ transform each other under the 
$Z_4$ twist $\theta$.
Thus, the $Z_4$ invariant states are written by \cite{Kobayashi:1990mc}
\begin{equation}
\sigma_{\theta^2,(0,0)},\qquad \sigma_{\theta^2,+}, \qquad \sigma_{\theta^2,(1,1)},
\end{equation}
while $\sigma_{\theta^2,-}$ transforms to $-\sigma_{\theta^2,-}$ under the $Z_4$ twist,
where 
\begin{equation}
\sigma_{\theta^2,\pm} =  \frac{1}{\sqrt 2} \left( \sigma_{\theta^2,(1,0)} \pm \sigma_{\theta^2,(0,1)} \right).
\end{equation}

Similarly, we can obtain the fixed points on $T^2/Z_6$.
There is a fixed point $(\theta,0)$ for the $Z_6$ twist $\theta$, and 
 a single twisted state $\sigma_{\theta,0}$.
The second twisted sector has three fixed points $(\theta^2,m)$ ($m=0,1,2$), which correspond to the three fixed points on 
the $T^2/Z_3$ orbifold.
The two fixed points $(\theta^2,1)$ and $(\theta^2,2)$ transform each other by the $Z_6$ twist, 
while $(\theta^2,0)$ is invariant.
Thus, we can write the $Z_6$-invariant $\theta^2$-twisted states by
\begin{equation}
\sigma_{\theta^2,0},\qquad \sigma_{\theta^2,+}, 
\end{equation}
while $\sigma_{\theta^2,-}$ transforms to $-\sigma_{\theta^2,-}$ under the $Z_6$ twist,
where 
\begin{equation}
\sigma_{\theta^2,\pm} =  \frac{1}{\sqrt 2} \left( \sigma_{\theta^2,1} \pm \sigma_{\theta^2,2} \right).
\end{equation}
The third twisted sector has four fixed points, which correspond to the fixed points on $T^2/Z_2$, 
and the corresponding $\theta^3$ twisted states.
Their linear combinations are $Z_6$ eigenstates similar to the second twisted states.
Since the first twisted sector has the single fixed point and twisted state, 
the modular symmetry as well as non-Abelian discrete flavor symmetry is rather trivial.
We do not discuss the $T^2/Z_6$ orbifold itself.

\subsection{Modular symmetry}

In Ref.~\cite{Lauer:1989ax}, modular symmetry in heterotic string theory on orbifolds was studied in detail.
Here we use those results.

\subsubsection{$T^2/Z_4$ orbifold}

The $S$ and $T$ transformations are represented by the first twisted sectors of $T^2/Z_4$ orbifold as \cite{Lauer:1989ax},
\begin{eqnarray}
& & \left( \begin{array}{c}
 \sigma_{\theta,0} \\ 
\sigma_{\theta,1} 
\end{array} \right) \longrightarrow S_{Z_4}\left( \begin{array}{c}
 \sigma_{\theta,0} \\ 
\sigma_{\theta,1}   
\end{array} \right), \qquad S_{Z_4} = \frac{1}{\sqrt 2}\left(
\begin{array}{cc}
1  & 1 \\
1  &  -1
\end{array}
\right), \nonumber \\
& & \left( \begin{array}{c}
 \sigma_{\theta,0} \\ 
\sigma_{\theta,1} 
\end{array} \right) \longrightarrow T_{Z_4}\left( \begin{array}{c}
 \sigma_{\theta,0} \\ 
\sigma_{\theta,1} 
\end{array} \right), \qquad T_{Z_4} = \left(
\begin{array}{cc}
1  & 0 \\
0  &  i
\end{array}
\right).
\end{eqnarray} 
These are exactly the same as representations of $S_{(2)}$ and $T_{(2)}$ on two-zero modes, 
$\psi^{0,2}$ and $\psi^{1,2}$ in the magnetized model with magnetic flux $M=2$.
Hence, the twisted sectors on the $T^2/Z_4$ orbifold has the same behavior of modular symmetry 
as the magnetized model with magnetic flux $M=2$.
Indeed, the twisted sectors have the $D_4$ flavor symmetry and two twisted states, 
$\sigma_{\theta,0}$ and $\sigma_{\theta,1}$ correspond to the $D_4$ doublet 
 \cite{Kobayashi:2006wq}.
The whole flavor symmetry of the $T^2/Z_4$ orbifold model is slightly larger than $D_4$.
(See Appendix \ref{sec:app-2}.)
The $T^2/Z_4$ orbifold model has the $Z_4$ symmetry, which transforms the first twisted sector,
\begin{equation}
\label{eq:Z4}
\sigma_{\theta,m} \longrightarrow e^{\pi i/2} \sigma_{\theta,m}, 
\end{equation}
for $m=0,1$ and the second twisted sector,
\begin{equation}
\sigma_{\theta^2,(m,n)} \longrightarrow e^{\pi i} \sigma_{\theta^2,(m,n)},
\end{equation}
for $m,n=0,1$.
The above $Z_4$ transformation (\ref{eq:Z4}) is nothing but $(S_{Z_4}T_{Z_4})^6$ 
as clearly seen from Eq.~(\ref{eq:ST-M=2}).
Thus, the whole flavor symmetry originates from the modular symmetry.

The second twisted sectors correspond to $D_4$ singlets, ${\bf 1}_{\pm 1, \pm}$ \cite{Kobayashi:2006wq} as 
\begin{eqnarray}
{\bf 1}_{+\pm}: \sigma_{\theta^2,(0,0)} \pm \sigma_{\theta^2,(1,1)}, \qquad 
{\bf 1}_{-\pm}: \sigma_{\theta^2,\pm},
\end{eqnarray}
up to coefficients.
Compared with the results in section \ref{sec:D4}, the $D_4$ behavior of the second twisted states 
correspond to one of the zero-modes $\psi^{m,4}$ with magnetic flux $M=4$.
Their correspondence can be written as 
\begin{eqnarray}
& & \sigma_{\theta^2,(0,0)} \sim \psi^{0,4}, \qquad \sigma_{\theta^2,(1,1)} \sim \psi^{2,4}, \nonumber \\
& &  \sigma_{\theta^2,(1,0)} \sim \psi^{1,4}, \qquad \sigma_{\theta^2,(1,0)} \sim \psi^{3,4}.
\end{eqnarray} 

The above correspondence can also been seen from the Yukawa couplings.
By use of operator product expansion, we obtain the following relations \cite{Lauer:1989ax},
\begin{eqnarray}
& & \sigma_{\theta,0} \sigma_{\theta,0} \sim Y_{0,0} \left( \sigma_{\theta^2,(0,0)} +  \sigma_{\theta^2,(1,1)}\right) , \nonumber \\
& & \sigma_{\theta,1} \sigma_{\theta,1} \sim Y_{1,1} \left( \sigma_{\theta^2,(0,0)} +  \sigma_{\theta^2,(1,1)} \right),  \\
& &  \sigma_{\theta,0} \sigma_{\theta,1} + \sigma_{\theta,1} \sigma_{\theta,0}\sim Y_{0,1} \sigma_{\theta^2,+}  \nonumber 
\end{eqnarray}
up to constants.
The second twisted state $\sigma_{\theta^2,-}$ can not couple with the first twisted sectors.
Using results in Ref.~\cite{Lauer:1989ax}, it is found that 
\begin{equation}
(T_{Z_4})^2 \left(
\begin{array}{c}
Y_{0,0} \\ Y_{1,1} \\ Y_{0,1}
\end{array} \right) = \left(
\begin{array}{ccc}
1 & 0 & 0 \\
0 & 1 & 0 \\
0 & 0 & -1 
\end{array}\right)
\left(
\begin{array}{c}
Y_{0,0} \\ Y_{1,1} \\ Y_{0,1}
\end{array} \right) .
\end{equation}
This is the same as behavior of the Yukawa couplings under $T^2$ studied in section \ref{sec:D4}.

\subsubsection{$T^2/Z_2$ orbifold}

Here, let us study the $T^2/Z_2$ orbifold in a way to similar to the previous section on the $T^2/Z_4$.
The $S$ transformation is represented by the four twisted states on the $T^2/Z_2$ orbifold \cite{Lauer:1989ax},
\begin{equation}
 \left( \begin{array}{c}
 \sigma_{\theta,(0,0)} \\ 
\sigma_{\theta,(0,1)}  \\
 \sigma_{\theta,(1,0)} \\ 
\sigma_{\theta,(1,1)}  
\end{array} \right) \longrightarrow S_{Z_2}\left( \begin{array}{c}
 \sigma_{\theta,(0,0)} \\ 
\sigma_{\theta,(0,1)}  \\
 \sigma_{\theta,(1,0)} \\ 
\sigma_{\theta,(1,1)}  
\end{array} \right), \qquad S_{Z_2} = \frac{1}{ 2}\left(
\begin{array}{cccc}
1  & 1 & 1 & 1\\
1  & -1 & 1 & -1 \\
1 & 1 & -1 & -1 \\
1  &  -1 & -1 & 1
\end{array}
\right).
\end{equation}
Also the $T$ transformation is represented as
\begin{equation}
 \left( \begin{array}{c}
 \sigma_{\theta,(0,0)} \\ 
\sigma_{\theta,(0,1)}  \\
 \sigma_{\theta,(1,0)} \\ 
\sigma_{\theta,(1,1)}  
\end{array} \right) \longrightarrow T_{Z_2}\left( \begin{array}{c}
 \sigma_{\theta,(0,0)} \\ 
\sigma_{\theta,(0,1)}  \\
 \sigma_{\theta,(1,0)} \\ 
\sigma_{\theta,(1,1)}  
\end{array} \right), \qquad T_{Z_2} = \left(
\begin{array}{cccc}
1  & 0 & 0 & 0\\
0  & -1 & 0 & 0 \\
0 & 0 & -1 & 0 \\
0  &  0 & 0 & -1
\end{array}
\right).
\end{equation}
The representation $S_{Z_2}$ is similar to $S_{Z_4}$ and $S_{(2)}$.
Indeed, we find that $S_{Z_2} = S_{(2)} \otimes S_{(2)}$.
However, the representation $T_{Z_2}$ is different from $T_{Z_4}$ and $T_{(2)}$.

The matrices $S_{Z_2}$ and $T_{Z_2}$ satisfy the following relations,
\begin{equation}
(S_{Z_2})^2=(T_{Z_2})^2=(S_{Z_2}T_{Z_2})^6=\mathbb{I}.
\end{equation}
These correspond to the $D_6$.
Indeed, the order of closed algebra including $S_{Z_2}$ and $T_{Z_2}$ is equal to 12.
At any rate, these matrices are reducible.
We change the basis in order to obtain irreducible representations, 
\begin{equation}
\left(
\begin{array}{c}
\sigma_1 \\
\sigma_2 \\
\sigma_3 \\
\sigma_4
\end{array}\right)=\left(
\begin{array}{cccc}
1 & 0 & 0& 0\\
0 & \frac{1}{\sqrt 3} & \frac{1}{\sqrt 3} & \frac{1}{\sqrt 3} \\
0 &  \frac{1}{\sqrt 2} & \frac{-1}{\sqrt 3} & 0 \\
0 & \frac{1}{\sqrt 6} & \frac{1}{\sqrt 6} & \frac{-2}{\sqrt 6} 
\end{array}\right)
\left(
\begin{array}{c}
\sigma_{\theta,(0,0)} \\
\sigma_{\theta,(1,0)} \\
\sigma_{\theta,(0,1)} \\
\sigma_{\theta,(1,1)}
\end{array}\right).
\end{equation}
Then, $\sigma_1$ and $\sigma_2$ correspond to the $D_6$ doublet, while 
$\sigma_3$ and $\sigma_4$ correspond to the $D_6$ singlets.
For example, $S_{Z_2}T_{Z_2}$ and $T_{Z_2}$ are represented by 
\begin{eqnarray}
& & S_{Z_2}T_{Z_2} 
 \left(
\begin{array}{c}
\sigma_1 \\
\sigma_2 \\
\sigma_3 \\
\sigma_4
\end{array}\right)=\left(
\begin{array}{cccc}
\cos (2\pi/6) & -\sin(2\pi/6) & 0 & 0 \\
\sin(2\pi /6) & \cos(2 \pi /6) & 0 & 0 \\
0 & 0& 1 & 0 \\
0 & 0& 0 & -1
\end{array}\right)
 \left(
\begin{array}{c}
\sigma_1 \\
\sigma_2 \\
\sigma_3 \\
\sigma_4
\end{array}\right),   \nonumber \\
& & T_{Z_2} 
 \left(
\begin{array}{c}
\sigma_1 \\
\sigma_2 \\
\sigma_3 \\
\sigma_4
\end{array}\right)=\left(
\begin{array}{cccc}
1 & 0 & 0 & 0 \\
0 & -1 & 0 & 0 \\
0 & 0& -1 & 0 \\
0 & 0& 0 & -1
\end{array}\right)
 \left(
\begin{array}{c}
\sigma_1 \\
\sigma_2 \\
\sigma_3 \\
\sigma_4
\end{array}\right).
\end{eqnarray}
It is found that $\sigma_{3}$ and $\sigma_4$ correspond to 
${\bf 1}_{--}$ and ${\bf 1}_{-+}$.

The twisted sector on the $T^2/Z_2$ orbifold has the flavor symmetry $(D_4 \times D_4)/Z_2$.
However, this flavor symmetry seems independent of the above $D_6$, 
because they do not include any common elements.
The twisted sector on the $S^1/Z_2$ orbifold has the flavor symmetry $D_4$.
The flavor symmetry of $T^2/Z_2$ orbifold is obtained as a kind of product, $D_4 \times D_4$, 
although two $D_4$ groups have a common $Z_2$ element.
Thus, the flavor symmetry of $T^2/Z_2$ originates from the product of symmetries of the one-dimensional orbifold.
On the other hand, the modular symmetry appears in two or more dimensions, but not in one dimension.
Hence, these symmetries would be independent.
When we include the above $D_6$ as low-energy effective field theory in addition to the flavor symmetry 
$(D_4 \times D_4)/Z_2$, low-energy effective field theory would have larger symmetry including $D_6$ and 
$(D_4 \times D_4)/Z_2$, although Yukawa couplings as well as higher order couplings transform 
non-trivially under $D_6$.

\subsubsection{$T^2/Z_3$ orbifold}

The $S$ and $T$ transformations are represented by the first twisted sectors of $T^2/Z_3$ orbifold as \cite{Lauer:1989ax},
\begin{eqnarray}
& & \left( \begin{array}{c}
 \sigma_{\theta,0} \\ 
\sigma_{\theta,1}   \\
\sigma_{\theta,2}
\end{array} \right) \longrightarrow S_{Z_3}\left( \begin{array}{c}
 \sigma_{\theta,0} \\ 
\sigma_{\theta,1}   \\
\sigma_{\theta,2} 
\end{array} \right), \qquad S_{Z_3} = \frac{1}{\sqrt 3}\left(
\begin{array}{ccc}
1  & 1 & 1\\
1  &  e^{2\pi i/3} & e^{-2\pi i/3} \\
1  &  e^{-2\pi i/3} & e^{2\pi i/3}
\end{array}
\right), \nonumber \\
& & \left( \begin{array}{c}
 \sigma_{\theta,0} \\ 
\sigma_{\theta,1} \\
\sigma_{\theta,2} 
\end{array} \right) \longrightarrow T_{Z_3}\left( \begin{array}{c}
 \sigma_{\theta,0} \\ 
\sigma_{\theta,1} \\
\sigma_{\theta,2} 
\end{array} \right), \qquad T_{Z_3} = \left(
\begin{array}{ccc}
1  & 0 & 0\\
0  &  e^{2\pi i/3} & 0 \\
0 & 0 & e^{2\pi i/3} 
\end{array}
\right).
\end{eqnarray} 
These forms look similar to $S$ and $T$ transformations in magnetized models (\ref{eq:magne-S}) and (\ref{eq:magne-T}).
Indeed, they correspond to submatrices of $S_{(6)}$ and $T_{(6)}$ in the magnetized models with 
the magnetic flux $M=6$.
Alternatively, in Ref.~\cite{Lerche:1989cs} the following $S$ and $T$ representations were studied\footnote{
See also Ref.~\cite{Ferrara:1989qb}.} 
\begin{eqnarray}
& &  S'_{Z_3} = -\frac{i}{\sqrt 3}\left(
\begin{array}{ccc}
1  & 1 & 1\\
1  &  e^{2\pi i/3} & e^{-2\pi i/3} \\
1  &  e^{-2\pi i/3} & e^{2\pi i/3}
\end{array}
\right), \qquad T'_{Z_3} = \left(
\begin{array}{ccc}
e^{2\pi i/3}  & 0 & 0\\
0  &  1 & 0 \\
0 & 0 & 1
\end{array}
\right).
\end{eqnarray}

At any rate, the above representations are reducible representations.
Thus, we use the flowing basis,
\begin{equation}
\left( 
\begin{array}{c}
\sigma_+ \\ \sigma_0 \\ \sigma_-
\end{array}
\right),  
\end{equation}
where $\sigma_\pm = (\sigma_1 \pm \sigma_-)/\sqrt 2$.
The $(\sigma_+,\sigma_0)$ is a doublet, while $\sigma_-$ is a singlet.
The former corresponds to the $Z_6$ invariant states among the $\theta^2$ twisted sector 
on the $T^2/Z_6$ orbifold.
Similarly, $\sigma_-$ is the $\theta^2$ twisted state, which transforms 
$\sigma_- \rightarrow -\sigma_-$ under the $Z_6$ twist.
Alternatively, we can say that the doublet $(\sigma_+,\sigma_0)$ corresponds to $Z_2$ even states and 
the singlet $\sigma_-$ is the $Z_2$ odd states, where the $Z_2$ means the $\pi$ rotation of the lattice vectors, 
$(\alpha_1,\alpha_2) \rightarrow (-\alpha_1,-\alpha_2)$. 
This point is similar to the aspect in magnetized D-brane models, where 
irreducible representations correspond to the $T^2/Z_2$ orbifold basis.
Also, note that the first twisted states of the $T^2/Z_4$ orbifold 
correspond already to the $Z_2$-invariant basis.

For example, we represent $S'_{Z_3}$ and $T'_{Z_3}$ on the above basis \cite{Lerche:1989cs} ,
\begin{equation}
S'_{Z_3} = \frac{i}{\sqrt{3}}\left(
\begin{array}{cc}
1 & \sqrt 2  \\
\sqrt 2 & -1
\end{array}
\right), \qquad T'_{Z_3} = \left(
\begin{array}{ccc}
e^{2 \pi i/3} &  0 \\
0 & 1 
\end{array} \right) ,
\end{equation}
on the doublet  $(\sigma_+,\sigma_0)^T$, while $\sigma_-$ is the trivial singlet.
Here, we define 
\begin{equation}
Z=\left(
\begin{array}{cc}
-1 & 0 \\
0 & -1
\end{array}
\right), \qquad \tilde T_{Z_3} = ZT'_{Z_3}.
\end{equation}
Then, they satisfy the following algebraic relations \cite{Lerche:1989cs,Ferrara:1989qb},
\begin{equation}
(S'_{Z_3})^2=(\tilde T_{Z_3})^3=(S'_{Z_3}\tilde T_{Z_3})^3=Z,\qquad Z^2=\mathbb{I}.
\end{equation}
This group is the so-called $T'$, which is the binary extension of $A_4 = T$.

The non-Abelian discrete flavor symmetry on the $T^2/Z_3$ orbifold is 
$\Delta(54)$, and the three twsisted states correspond to the triplet of 
$\Delta(54)$.
Thus, this modular symmetry seems independent of the $\Delta(54)$ flavor symmetry.

Two representations are related as 
\begin{equation}
S'_{Z_3} = -i S_{Z_3}, \qquad T'_{Z_3} = e^{2\pi i /3}(T_{Z_3})^{-1}.
\end{equation}
When we change phases of $S$, $T$ and $ST$, the group such as $(Z_N \times Z_M) \rtimes H$ 
in sections \ref{sec:magne-D} \ref{sec:Het} and  would change to 
$(Z_{N'} \times Z_{M'}) \rtimes H$.


\section{Conclusion}
\label{sec:conclusion}

We have studied the modular symmetries in magnetized D-brane models.
Representations due to zero-modes on $T^2$ are reducible except the models with the magnetic flux $M=2$.
Irreducible representations are provided by zero-modes on the $T^2/Z_2$, i.e. $Z_2$ even states and 
odd states.
It is reasonable because $(ST)^3$ transforms the lattice vectors $(\alpha_1,\alpha_2)$ to $(-\alpha_1,-\alpha_2)$. 
The orders of modular groups are large, and in general, they include the $Z_8$ symmetry as the center.
The $D_4$ flavor symmetry is a subgroup of the modular group, which is represented in 
the models with the magnetic flux $M=2$.
The system including zero-modes with $M=2$, $M=4$ and larger even $M$, also includes 
the $D_4$ flavor symmetry, when we define transformations of couplings in a proper way.

We have also studied the modular symmetries in hetetrotic orbifold models.
The heterotic model on the $T^2/Z_4$ has exactly the same representation as the magnetized model 
with $M=2$, and the modular symmetry includes the $D_4$ flavor symmetry.
The representation due to the twisted states on the $T^2/Z_3$ orbifold is reducible, similar to 
representations due to zero-modes in magnetized D-brane models on $T^2$.
Their irreducible representations correspond to $Z_2$ even and odd states, 
similar to those in magnetized D-brane models.
Thus, the $\Delta(54)$ flavor symmetry seems independent of the modular symmetry 
in the $T^2/Z_3$ orbifold models.
Note that the first twisted states on the $T^2/Z_4$ are $Z_2$-invariant states.
In this sense, we find that the modular symmetry is the symmetry 
on the $Z_2$ orbifold in both heterotic orbifold models and magnetized D-brane models.
The symmetries, which remain under the $Z_2$ twist, can be realized 
as the modular symmetry.

We have set vanishing Wilson lines.
It would be interesting to extend our analysis to 
 magnetized D-brane models with discrete Wilson lines on orbifolds \cite{Abe:2013bca}.
It would be also interesting to extend our analysis on zero-modes to higher Kaluza-Klein modes \cite{Hamada:2012wj}.

Four-dimensional low energy-effective field theory is modular invariant \cite{Ferrara:1989bc,Ferrara:1990ei,Cvetic:1991qm}.
Anomalies of the modular symmetry  were studied \cite{Derendinger:1991hq,Ibanez:1992hc}, 
and they have important aspects \cite{Ibanez:1991zv,Kawabe:1994mj,Kobayashi:2016ovu}.
The non-Abelian flavor symmetries such as $D_4$ can be anomalous.
(See for anomalies of non-Abelian discrete symmetries, e.g. \cite{Araki:2008ek,Ishimori:2010au,Talbert:2018nkq}. )
In certain models, the modular symmetries are related with the non-Abelian flavor symmetry $D_4$.
It would be interesting to study their anomaly relations.

We also give a comment on phenomenological application.
Recently, the mixing angles in the lepton sector were studied in the models, 
whose flavor symmetries are congruence subgroups, $\Gamma(N)$ \cite{Feruglio:2017spp,Kobayashi:2018vbk}.
In those models, the couplings are non-trivial representations of $\Gamma(N)$ 
and modular functions.
Our models show massless modes represent larger finite groups.
It would be interesting to apply our results to derive realistic lepton mass matrices 
as well as quark mass matrices.


\section*{Acknowledgments}
T.~K. was is supported in part by MEXT KAKENHI Grant Number JP17H05395 and 
JSP KAKENHI Grant Number JP26247042.

%


\appendix

\section{Non-Abelian discrete flavor symmetry in magnetized D-brane models}
\label{sec:app-1}

In this Appendix, we give a brief review on non-Abelian discrete flavor 
symmetries in magnetized D-brane models \cite{Abe:2009vi}.

As mentioned in section \ref{sec:zero-mode}, the Yukawa coulings as well as higher order couplings have 
the coupling selection rule (\ref{eq:selection}).
That is, we can define $Z_g$ charges for zero-modes.
Such $Z_g$ transformation is represented on $\psi^{i,M=g}$ by 
\begin{equation}
Z = \left( 
\begin{array}{ccccc}
1 & & & & \\ 
 &\rho & & & \\
 & & \rho^2& & \\
 & & & \ddots & \\
 & & & & \rho^{g-1}
\end{array} \right),
\end{equation}
where $\rho = e^{2 \pi i /g}$.
Furthermore, their effective field theory has the following permutation symmetry,
\begin{equation}
\psi^{i,g} \to \psi^{i+1, g},
\end{equation}
and such permutation can be represented by 
\begin{equation}
\label{eq:nonableC}
C =\left( 
\begin{array}{cccccc}
0 & 1 & 0 & 0 & \cdots & 0 \\ 
0 & 0 & 1 & 0 & \cdots & 0 \\
  &    &    &   & \ddots & \\
1 & 0 & 0 & 0 & \cdots & 0
\end{array} \right).
\end{equation}
This is another $Z_g^C$ symmetry.
However, these two generators do not commute each other, 
\begin{equation}
CZ = \rho ZC.
\end{equation}
Thus, the flavor symmetry corresponds to the closed algebra 
including $Z$ and $C$.
Its diagonal elements are given by 
$Z^mZ'^n$, i.e. $Z_g \times Z'_g$ where 
\begin{equation}
Z' = \left(
\begin{array}{ccc}
\rho & & \\
 & \ddots & \\
 & & \rho
\end{array} \right),
\end{equation}
and the full group corresponds to 
$(Z_g \times Z'_g) \rtimes Z_g^C$.

Furthermore, the zero-modes $\psi^{i,M=gn}$ with the magnetic flux $M=gn$ also 
represent $(Z_g \times Z'_g) \rtimes Z_g^C$.
The zero-modes, $\psi^{i,M=gn}$ have $Z_g$ charges (mod $g$).
Under $C$, they transform as 
\begin{equation}
\psi^{i,M=gn} \to \psi^{i + n, M=gn} .
\end{equation}

For example, the model with $g=2$ has the $D_4$ flavor symmetry.
The zero-modes, 
\begin{equation}
\left(
\begin{array}{c}
\psi^{0,2} \\ \psi^{1,2}
\end{array}
\right),
\end{equation}
correspond to the $D_4$ doublet ${\bf 2}$, 
where eight $D_4$ elements are represented by
\begin{equation}\label{eq:D4-8}
\pm\left(
\begin{array}{cc}
1 & 0 \\
0 & 1
\end{array}
\right), \qquad  \pm \left(
\begin{array}{cc}
0 & 1 \\
1 & 0
\end{array}
\right), \qquad  \pm \left(
\begin{array}{cc}
0 & 1 \\
-1 & 0
\end{array}
\right), \qquad  \pm \left(
\begin{array}{cc}
1 & 0 \\
0 & -1
\end{array}
\right).
\end{equation}
In addition, when the model has the zero-modes $\psi^{i,4}$ $(i=0,1,2,3)$, 
the zero-modes, $\psi^{0,4}$ and $\psi^{2,4}$ ( $\psi^{1,4}$ and $\psi^{3,4}$)  
transform each other under $C$, and they have $Z_2$ charge even (odd).
Thus, $\psi^{0,4}\pm\psi^{2,4}$ correspond to ${\bf 1}_{+\pm}$ of $D_4$ representations, 
while $\psi^{1,4}\pm\psi^{3,4}$ correspond to ${\bf 1}_{-\pm}$.
Furthermore, among the zero-modes $\psi^{i,6}$ $(i=0,1,2,3,4,5)$, 
the zero-modes $\psi^{i,6}$ and $\psi^{i+3,6}$ transform each other under $C$.
Hence, three pairs of zero-modes,  
\begin{equation}
\left(
\begin{array}{c}
\psi^{0,6} \\ \psi^{3,6}
\end{array} \right), \qquad 
\left(
\begin{array}{c}
\psi^{1,6} \\ \psi^{4,6}
\end{array} \right), \qquad
\left(
\begin{array}{c}
\psi^{2,6} \\ \psi^{5,6}
\end{array} \right), \
\end{equation}
correspond to three $D_4$ doublets.
These results are shown in Table \ref{tab:D4}.

\begin{table}[ht]   
\begin{center}
\begin{tabular}{|c|c|}  \hline
Magnetic flux $M$ &   $D_4$ representations \\  \hline \hline
2 & ${\bf 2}$  \\
4 & ${\bf 1}_{++}$, ${\bf 1}_{+-}$, ${\bf 1}_{-+}$, ${\bf 1}_{--}$  \\
6 & 3 $\times {\bf 2}$    \\ \hline
\end{tabular}
\caption{$D_4$ representation}   \label{tab:D4}
\end{center}
\end{table}

\section{Non-Abelian discrete flavor symmetry in heterotic orbifold models}
\label{sec:app-2}

Here, we give a brief review on non-Abelian discrete flavor symmetries 
in heteotic orbifold models \cite{Kobayashi:2006wq}.

The twisted string $x^i$ on the orbifold satisfy the following boundary condition:
\begin{equation}
\label{eq:string-BC}
x^i(\sigma = 2 \pi) = (\theta^n x(\sigma = 0))^i + \sum_k m_k \alpha_k^i,
\end{equation}
similar to Eq.~(\ref{eq:fp}).
Thus, the twisted string can be characterized by the space group element 
$g= (\theta^{n}, \sum_k m_k \alpha_k^i)$.
The product of the two space group elements $(\theta^{n_1},v_1)$ and $(\theta^{n_2},v_2)$
is computed as
\begin{equation}
(\theta^{n_1},v_1)(\theta^{n_2},v_2) = (\theta^{n_1}\theta^{n_2},v_1 + \theta^{n_1}v_2).
\end{equation}
The space group element $g$ belongs to the same conjugacy class as $hgh^{-1}$, 
where $h$ is any space group element on the same orbifold.

Now, let us consider the couplings among twisted strings corresponding to 
space group elements $(\theta^{n_k},v_k)$.
Their couplings are allowed by the space group invariance if the following 
condition:
\begin{equation}
\prod_k (\theta^{n_k},v_k) = (1,0),
\end{equation}
is satisfied up to the conjugacy class.
That includes the point group selection rule, $\prod_k \theta^{n_k} = 1$, 
which is the $Z_N$ invariance on the $Z_N$ orbifold.
We can define discrete Abelian symmetries from the space group invariance as well as 
the point group invariance.
These symmetries together with geometrical symmetries of orbifolds become 
non-Abelian discrete flavor symmetries in heterotic orbifold models.
We show them explicitly on concrete orbifolds.

\subsection{$S^1/Z_2$ orbifold}

The $S^1/Z_2$ orbifold has two fixed points, which are denoted by the space group elements, 
$(\theta,m\alpha)$ with $m=0,1$, where $\alpha$ is the lattice vector.
In short, we denote them by $(\theta,m)$ and the corresponding twisted states are 
denoted by $\sigma_{(\theta,m)}$.
These states transform
\begin{equation}
\label{eq:Z2-twist}
\left(
\begin{array}{c}
\sigma_{\theta,0} \\ \sigma_{\theta,1}
\end{array}\right) \longrightarrow 
\left(
\begin{array}{cc}
-1 & 0 \\
0  & -1
\end{array} \right)
\left(
\begin{array}{c}
\sigma_{\theta,0} \\ \sigma_{\theta,1}
\end{array}\right) ,
\end{equation}
under the $Z_2$ twist.
In addition, the space group invariance requires 
$\sum_k m_k =0$ (mod 2) for the couplings corresponding to 
the product of the space group elements $\prod_k (\theta,m_k)$ with $m_k=0,1$.
Hence, we can define another $Z_2$ symmetry, under which 
$\sigma_{(\theta,0)}$ is even, while  $\sigma_{(\theta,1)}$ is odd.
That is, another $Z_2$ transformation can be written by 
\begin{equation}
\label{eq:Z2-space}
\left(
\begin{array}{c}
\sigma_{\theta,0} \\ \sigma_{\theta,1}
\end{array}\right) \longrightarrow 
\left(
\begin{array}{cc}
1 & 0 \\
0  & -1
\end{array} \right)
\left(
\begin{array}{c}
\sigma_{\theta,0} \\ \sigma_{\theta,1}
\end{array}\right) .
\end{equation}
Furthermore, there is the geometrical permutation symmetry, which exchange two fixed points each other.
Such a permutation is represented by 
\begin{equation}
\label{eq:Z2-perm}
\left(
\begin{array}{c}
\sigma_{\theta,0} \\ \sigma_{\theta,1}
\end{array}\right) \longrightarrow 
\left(
\begin{array}{cc}
0 & 1 \\
1  & 0
\end{array} \right)
\left(
\begin{array}{c}
\sigma_{\theta,0} \\ \sigma_{\theta,1}
\end{array}\right) .
\end{equation}
The closed algebra including Eqs.(\ref{eq:Z2-twist}), (\ref{eq:Z2-space}) and (\ref{eq:Z2-perm}) 
is $D_4 \simeq (Z_2 \times Z_2) \rtimes Z_2$.

\subsection{$T^2/Z_3$ orbifold}

As shown in Section \ref{sec:Het}, the $T^2/Z_3$ orbifold has three fixed points denoted by 
$(\theta,m)$ with $m=0,1,2$, and the corresponding twisted states are denote by 
$\sigma_{(\theta,m)}$.
The $Z_3$ twist transforms 
\begin{equation}
\label{eq:Z3-twist}
\left(
\begin{array}{c}
\sigma_{\theta,0} \\ \sigma_{\theta,1} \\ \sigma_{\theta,2}
\end{array}\right) \longrightarrow 
\left(
\begin{array}{ccc}
e^{2\pi i/3} & 0 & 0 \\
0  & e^{2 \pi i/3} & 0 \\
0  & 0 &  e^{2 \pi i/3}
\end{array} \right)
\left(
\begin{array}{c}
\sigma_{\theta,0} \\ \sigma_{\theta,1} \\ \sigma_{\theta,2}
\end{array}\right) .
\end{equation}

The space group invariance requires 
$\sum_k m_k =0$ (mod 3) for the couplings corresponding to 
the product of the space group elements $\prod_k (\theta,m_k)$ with $m_k=0,1,2$.
Then, we can define another $Z_3$ symmetry, under which 
$\sigma_{(\theta,m)}$ transform 
\begin{equation}
\label{eq:Z3-space}
\left(
\begin{array}{c}
\sigma_{\theta,0} \\ \sigma_{\theta,1} \\ \sigma_{\theta,2}
\end{array}\right) \longrightarrow 
\left(
\begin{array}{ccc}
1 & 0 & 0 \\
0  & e^{2\pi i/3} & 0 \\
0 & 0& e^{2\pi i/3} 
\end{array} \right)
\left(
\begin{array}{c}
\sigma_{\theta,0} \\ \sigma_{\theta,1} \\ \sigma_{\theta,2}
\end{array}\right) .
\end{equation}
There is also the permutation symmetry of the three fixed points, 
that is, $S_3$.
Thus, the flavor symmetry is $\Delta(54) \simeq (Z_3 \times Z_3) \rtimes S_3$.

\subsection{$T^2/Z_4$ orbifold }

As shown in Section \ref{sec:Het}, the $T^2/Z_4$ orbifold has two $\theta$ fixed points denoted by 
$(\theta,m)$ with $m=0,1$, and the corresponding twisted states are denote by 
$\sigma_{(\theta,m)}$.
The $Z_4$ twist transforms 
\begin{equation}
\label{eq:Z4-twist}
\left(
\begin{array}{c}
\sigma_{\theta,0} \\ \sigma_{\theta,1} 
\end{array}\right) \longrightarrow 
\left(
\begin{array}{cc}
i  & 0 \\
0  & i 
\end{array} \right)
\left(
\begin{array}{c}
\sigma_{\theta,0} \\ \sigma_{\theta,1} 
\end{array}\right) .
\end{equation}

The space group invariance requires 
$\sum_k m_k =0$ (mod 2) for the couplings corresponding to 
the product of the space group elements $\prod_k (\theta,m_k)$ with $m_k=0,1$.
Then, we can define another $Z_2$ symmetry, under which 
$\sigma_{(\theta,m)}$ transform 
\begin{equation}
\label{eq:Z4-space}
\left(
\begin{array}{c}
\sigma_{\theta,0} \\ \sigma_{\theta,1} \\ \sigma_{\theta,2}
\end{array}\right) \longrightarrow 
\left(
\begin{array}{ccc}
1 & 0  \\
0  & -1 
\end{array} \right)
\left(
\begin{array}{c}
\sigma_{\theta,0} \\ \sigma_{\theta,1} 
\end{array}\right) .
\end{equation}
There is also the permutation symmetry of the two fixed points.
Thus, the flavor symmetry is almost the same as one on the $S^1/Z_2$ orbifold.
The difference is the $Z_4$ twist, although its squire is nothing but the $Z_2$ twist.
Hence, the flavor symmetry can be written as $(D_4 \times Z_4)/Z_2$.

\subsection{$T^2/Z_2$ orbifold}

As shown in Section \ref{sec:Het}, the $T^2/Z_4$ orbifold has two $\theta$ fixed points denoted by 
$(\theta,(m,n))$ with $m,n=0,1$, and the corresponding twisted states are denote by 
$\sigma_{\theta,(m,n)}$.
The space group invariance requires 
$\sum_k m_k =\sum_j n_j=0$ (mod 2) for the couplings corresponding to 
the product of the space group elements $\prod_k (\theta,(m_k,n_j))$ with $m_k,n_j=0,1$.
There are two independent permutation symmetries between $(\theta,(0,n))$ and $(\theta,(1,n))$, 
and $(\theta,(m,0))$ and $(\theta,(m,1))$.
Thus, this structure seems be a product of two one-dimensional orbifolds, $S^1/Z_2$.
However, the $Z_2$ twist is comment such as $\sigma_{\theta,(m,n)} \longrightarrow -\sigma_{\theta,(m,n)}$.
Thus, the flavor symmetry can be written by $(D_4 \times D_4)/Z_2$.



\end{document}